\documentstyle[twocolumn,prc,aps,epsfig]{revtex}

\input epsf
\newcommand{\be}{\begin{eqnarray}}
\newcommand{\ee}{\end{eqnarray}}

 \newcommand{\gsim}{\mathrel{\hbox{\rlap{\lower.55ex \hbox {$\sim$}}
                   \kern-.3em \raise.4ex \hbox{$>$}}}}
\newcommand{\lsim}{\mathrel{\hbox{\rlap{\lower.55ex \hbox {$\sim$}}
                   \kern-.3em \raise.4ex \hbox{$<$}}}}

\def\roughly#1{\mathrel{\raise.3ex\hbox{$#1$\kern-.75em%
\lower1ex\hbox{$\sim$}}}}
\def\lsim{\roughly<}
\def\gsim{\roughly>}

\renewcommand{\thefootnote}{\arabic{footnote}}
\begin{document}

\twocolumn[\hsize\textwidth\columnwidth\hsize\csname @twocolumnfalse\endcsname

\title{ Magnetic Component of Quark-Gluon Plasma is also a Liquid!}
\author {Jinfeng Liao and Edward Shuryak}
\address {Department of Physics and Astronomy, State University of New York,
Stony Brook, NY 11794}
\date{\today}
\maketitle
\begin{abstract}
The magnetic scenario recently suggested in \cite{Liao_ES_mono}
emphasizes the role of monopoles in strongly coupled quark-gluon
plasma (sQGP) near/above the deconfinement temperature, and
specifically predicts that they help reduce its viscosity by the
``magnetic bottle'' effect. Arguments for ``magnetic liquid'' in
1-2$T_c$ based on lattice results of monopole density were
provided in \cite{Chernodub}. Here we present results for
monopole-(anti)monopole correlation functions from the same
classical molecular dynamics simulations, which are found to be in
good agreement with lattice results in \cite{D'Alessandro:2007su}.
We show that the magnetic Coulomb coupling runs in the direction
$opposite$ to the electric one, as expected, and it is roughly
inverse of the asymptotic freedom formula for the electric one.
However, as $T$ decreases to $T_c$, the magnetic coupling never
gets too weak, with the plasma parameter always large enough
($\Gamma>1$). This nicely agrees with empirical evidences from
RHIC experiments, implying that magnetic objects cannot have large
mean free path and should also form a good liquid with low
viscosity.
\end{abstract}
\vspace{0.6in} ]
\newpage

\section{Running couplings}
Discussions of electric-magnetic duality appear in theoretical
literature regularly since Maxwell's time.
 Especially important for what follows
are: (i) the celebrated Dirac
quantization condition
 \cite{Dirac}, (ii) 't Hooft-Polyakov monopole solutions
\cite{'t Hooft-Polyakov}, (iii) ``dual
superconductor'' idea of confinement by 't Hooft and Mandelstam
 and (iv) Seiberg-Witten
solution of the  ${\cal N}=2$ SUSY gauge theory \cite{SeiWit},
 identifying properties and dynamical role of
magnetically charged objects in this setting.

Before we focus on our main subject -- monopole mutual
interactions -- let us first comment on other theoretical issues
related with situations in which both ``electric'' and
``magnetic'' particles are present at the same time.

For pure gauge fields electric/magnetic duality simply means
rewriting magnetic field $B$ as gradient and
electric field $E$ as a curl of a ``dual''
potential: however there are nontrivial questions
about  the sources (and boundary conditions).
The electric objects -- quarks -- are traditionally present in the
Lagrangian as Noether charges, while monopoles are solitonic
solutions carrying topological charges. Can ``magnetic''
formulation be consistently defined, interchanging their roles and
putting monopoles in the Lagrangian instead? Can even a situation
be found in which both formulations are similar? These ideas were
discussed starting from the famous paper \cite{Montonen-Olive}.

 Since {\em both description
should describe the same theory} serious issues of consistency
appear. At the quantum mechanics level the famous Dirac
\cite{Dirac} condition must be held, demanding basically that the
product of two couplings is fixed to an integer (put here to 1)
\be \label{Dirac}
\alpha_E*\alpha_M=1 \ee
So while
one of them may be small, the other must necessarily be large.

At the level of quantum field theory the Dirac condition elevates
into a requirement that two couplings must run into the {\em
opposite directions}
\begin{equation}
\label{Dirac_Run} \tilde{\beta}(\alpha)_{E} +
\tilde{\beta}(\alpha)_{M} = 0
\end{equation}
 where the beta functions are $\tilde{\beta}(\alpha)\equiv
\frac{\mu}{\alpha}\frac{d\alpha}{d\mu}=2\beta(g)/g$ for the
electric and magnetic couplings respectively, with $\beta(g)\equiv
\mu\frac{dg}{d\mu}$ being the usual beta function. This indeed is
what happens in Seiberg-Witten solution, in which electric
coupling is weak at large momenta due to asymptotic freedom,
 and magnetic is weak at small ones due to U(1) ``Landau pole''.

In our previous paper \cite{Liao_ES_mono} we discussed for sQGP
the important role of this generic feature. As it is known for 30
years, QGP  at $high$ $T$ can be described perturbatively, with
 e.g. small quark and gluon effective masses
 $M/T\sim \sqrt{\alpha_{electric}}<<1$ 
The monopoles in this case are heavy composites
which play a minor role, although
they are strongly interacting and form an interesting sub-sector
in which perturbative analysis is impossible.
However as $T$ goes down  and
 one approaches the deconfinement transition $T\rightarrow T_c$,
 the inverse is expected to happen: electrically charged particles -- quarks
and gluons -- are getting heavier and more strongly coupled. There
are strong evidences that both phenomena do happen: lattice data
show that quark (baryon) masses seem to be large in sQGP near
$T_c$ \cite{Liao_ES_susceptibilities}, while RHIC's ``perfect
liquid'' \cite{Shuryak_03,discovery_workshop} supports the idea of
strong coupling.

However at this point one may ask what happens with monopoles: as
$T\rightarrow T_c$ the same logic suggests that they must
 become lighter and more important. At some point
their masses (and roles) get comparable to that of the electric
objects, after which the tables are turned and their fortune
reversed. Electric objects gets strongly coupled and complicated
while monopoles gets lighter, proliferate and eventually take over
the bulk, expelling electric fields into the flux tube. As shown
in \cite{Liao_ES_tube}, this may happen in the plasma phase,
before confinement transition is reached. The corresponding phase
diagram was discussed in \cite{Liao_ES_mono}. Here are two
questions, on which will we be focused below: (i) Are there
evidences that the magnetic coupling does run in the opposite
direction? (ii) How small does the magnetic coupling become at
$T\rightarrow T_c$, and is a perturbative description of magnetic
plasma  possible?  As the reader will see, we will answer ``yes''
to the first and ``no'' to the second question.

(A digression about the most symmetric
 $\cal N$=4 SYM theory which is conformal. How do we know that its
 coupling does not run? One may calculate the first coefficient
of the beta function, and will indeed see that negative gauge
contribution is nicely cancelled by fermions and scalars. But
there are infinitely many coefficients, and one has to check them
all! An elegant way to prove the case is based on another
outstanding feature of the $\cal N$=4 SYM: this theory is (nearly)
$self-dual$ under electric-magnetic duality. As we discussed
above, the Dirac condition requires the product of electric and
magnetic couplings to be constant. But, as shown by Osborn
\cite{Osborn}, in Higgs case the multiplet of (lowest) magnetic
objects of the $\cal N$=4 SYM theory include 5 scalars,
 4 fermions (monopoles plus one gluino zero mode
occupied), plus 1 spin-1 particle (3 polarizations),
 exactly the same set of
states as in the original electric multiplet
(gluon-gluinoes-Higgses). Thus effective magnetic theory has the
$same$ Lagrangian as the original electric formulation and
 the same beta function. That would conflict
with requirement that both couplings run in the opposite
direction, unless they do not run at all!)

\section{Molecular dynamics with monopoles}

It has became apparent around 2003
\cite{Shuryak_03,discovery_workshop} that Quark-Gluon Plasma
 discovered at Relativistic Heavy Ion Collision (RHIC)
is the most ``perfect liquid'' known. Theorists have since been
taking the big challenge to explain the remarkable properties of
sQGP with diverse approaches, ranging from those borrowed from
classical plasmas to AdS/CFT duality: for recent reviews see
\cite{Shuryak_08}.

Molecular Dynamics (MD) simulations are proved to be powerful
tools widely used for studying the conventional EM plasma,
especially in the strongly coupled regime when the analytic
approaches are difficult. It provides detailed microscopic
 real time information about correlation functions
 and transport properties. It has recently been employed to
 simulate sQGP in \cite{Gelman,Liao_ES_mono}. The key
 classical Coulomb plasma parameter $\Gamma$ is defined by
\begin{equation} \label{plasma_gamma}
\Gamma \equiv \frac{\alpha_C \, / \, (\frac{3}{4\pi n})^{1/3}}{T}
\end{equation}
with $\alpha_C$ the Coulomb
 coupling between charges, $Q=E$ or $M$, and $n$ the
density of the corresponding charges. It means the ratio of the
potential energy (interaction) to the thermal kinetic energy; thus
a weakly coupled plasma has $\Gamma<<1$.  Plasmas with $\Gamma>1$
are known as
 strongly coupled, and for $\Gamma =1..\Gamma_c\sim 100$
 it is in the liquid
regime, becoming solid for $\Gamma > \Gamma_c$. Less
precisely defined ``glassy'' regime is  at $Gamma=10..\Gamma^c$
in between. The electric $\Gamma$
parameter in sQGP is believed to be $\Gamma_E \sim 3$ at
$T=(1-2)T_c$ \cite{Gelman}, so it corresponds to a liquid. The
value of the magnetic parameter  $\Gamma_M$ is the subject of this
Letter.

\begin{figure}[t]
   \epsfig{file=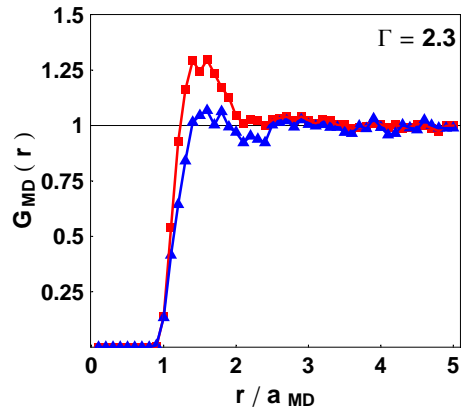,width=6.cm}
  \vspace{0.1in}
\caption{(color online)
 Monopole-antimonopole (boxes with red curve) and monopole-monopole (triangles with blue curve)
 correlators from MD simulations at $\Gamma=2.3$\protect\cite{Liao_ES_mono}.} \label{fig_gr_MD}
\end{figure}

In our previous paper \cite{Liao_ES_mono}
 we explored (to our knowledge,
for the first time) a novel plasmas made of a mixture of both
electric and magnetic charges. Using MD and standard Kubo formulae
we calculated its transport properties, i.e. the shear viscosity
and diffusion constant. We found that an equal 50\%-50\% mixture
has smallest viscosity (the shortest mean free path, if one want
to use such language) because of the ``magnetic bottle' effect.

Here we present an important property that was not discussed in
\cite{Liao_ES_mono}, namely the particle-particle equal-time
density-density correlator depending on the distance $r$:
\begin{equation} \label{correlator}
G_{ab}(r) \equiv \frac{ < \frac{1}{V}\sum_{i=1,N_a}\sum_{j=1,N_b}
\delta(|{\vec r}^{\, a}_i-{\vec r}^{\, b}_j|-r)>}{\rho_a \rho_b
4\pi r^2}
\end{equation}
where $a,b$ denotes two species of particles with total numbers
$N_a,N_b$ in a volume $V$ and $\rho_a,\rho_b$ their density. In
our MD simulation the monopole and anti-monopole numbers are both
set to be 250 inside a sphere with radius $10\, a_{MD}$. $a_{MD}$
is the length unit in MD and corresponds to about $0.2fm$ after
mapping to sQGP system: see \cite{Liao_ES_mono} for more details.
The correlator of the monopole-anti-monopole pair is particularly
instructive because it tells how strongly the monopoles are
correlated. The result is shown in Fig.\ref{fig_gr_MD}, for the
50\%-50\% electric-magnetic symmetric plasma with $\Gamma=2.3$. It
features a considerable nearest-neighbor peak in the
monopole-anti-monopole correlator, reaching around 1.25, with
hints to small secondary correlations: a typical $liquid$
behavior, see various examples in e.g. \cite{correlator_1}. The
same correlator for a weakly coupled gas will have a very weak
peak, barely above 1, while a solid will have strong multiple
peaks. The monopole-monopole correlator shows strong suppression
at short distance while stays very close to 1 at larger $r$: also
typical of Coulomb plasma. The reader shall recall that we do
$not$ consider here the usual Coulomb plasma due to presence of
equal number of electric particles in our mixture -- with which
monopoles interacts with a strong Lorentz force, leading to very
complicated motion -- so the result is far from trivial.

\begin{figure}[t]
   \epsfig{file=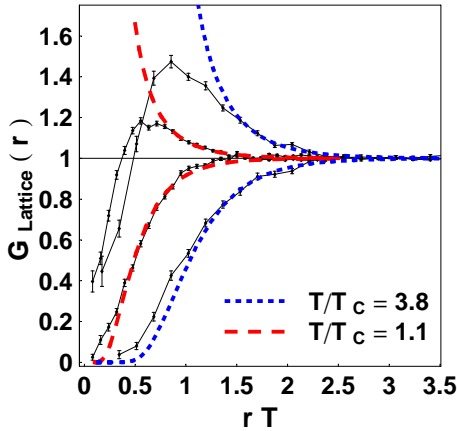,width=6.cm}
  \vspace{0.1in}
\caption{(color online) Monopole-antimonopole (the upper two
curves) and monopole-monopole (the lower two curves) correlators
at $1.1T_c$(red long dashed) and $3.8T_c$(blue dot dashed): points
with error bars are lattice data
\protect\cite{D'Alessandro:2007su}, the dashed lines are our fits
(see text). \label{fig_gr_fit}}
\end{figure}

\section{Lattice data and monopole correlations}

Lattice studies of monopoles have a long history which we would
not even attempt to summarize here (see e.g.
 \cite{Chernodub_2}). We will not discuss here the
properties of these monopoles  -- such as
masses or their interaction with
quarks/gluons or   ``Higgs field'' $A_0$ - to be addressed  in
\cite{RS} ). We will focus entirely on one aspect of these data,
related to the correlators (\ref{correlator}) and
$\Gamma$ parameter just discussed.
Those have been obtained recently in \cite{D'Alessandro:2007su},
for technical reasons for
the simplest
SU(2) pure gauge theory. Fig.\ref{fig_gr_fit} shows two sets
of their data for $T=1.1T_c$ and $T=3.8T_c$ for monopole-antimonopole
and for monopole-monopole correlations.

 Note that the curves are very similar to the
ones in our MD studies shown in previous section, especially
 the shape
and magnitude of the opposite-sign
peak. The most important feature
apparent from these (and other)
plots is that the correlation gets $stronger$
at the $higher$ $T$, confirming our expectation that the magnetic
component of sQGP gets stronger coupled at high $T$, $oppositely$
to the electric component.

To get a quantitative measure we have fitted them at large r (where
deviation from 1 is small and linearized screening
is correct) by the Debye formula
\begin{equation}\label{gr_large}
G_{ab}(r) \sim exp{ \bigg [\pm {\frac{\alpha_M e^{- r/R_d}}{rT}
} \bigg ]}
\end{equation}

\begin{figure}[t]
 \epsfig{file=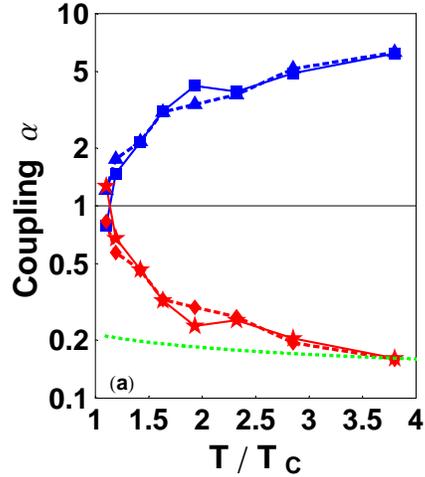,width=5.5cm}\vspace{0.2in}
 \epsfig{file=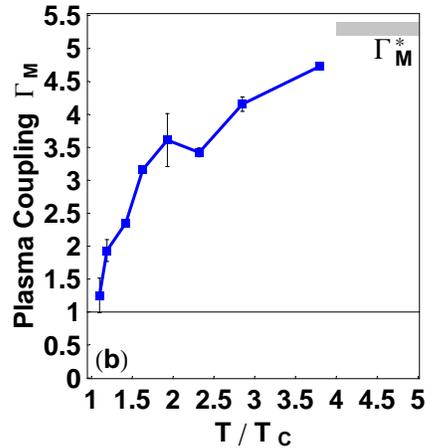,width=5.5cm}
  \vspace{0.1in}
\caption{(color online) (a) The magnetic coupling $\alpha_M$ (on
$Log_{10}$ scale) versus $T/T_c$ fitted from the
monopole-antimonopole (boxes with solid blue curve) and
monopole-monopole (triangles with dashed blue curve) correlators.
Their inverse, the corresponding $\alpha_E$ from the Dirac
condition, are shown as stars with solid red curve and diamonds
with dashed red curve respectively, together with an asymptotic
freedom (green dotted) curve (see text); (b) Effective magnetic
plasma coupling $\Gamma_M$
 versus $T/T_c$, with the gray band schematically showing
 high-T limit $\Gamma^{*}_M$ (see text).
\label{fig_running_coupling}}
\end{figure}

In (\ref{gr_large}) the positive sign in the exponent is for
monopole-antimonopole ($ab=+-$) while the negative for
monopole-monopole ($ab=++$), with $\alpha_M$ is the magnetic
coupling, and the $R_d$ magnetic screening radius. See the dashed
curves in Fig.\ref{fig_gr_fit}. Some details about the fitting for
each data set: we exclude the last few points that are rather flat
and marginally lower/higher than 1 for $G_{+-}$/$G_{++}$ which the
fitting formulae can not reach;
 the rule to determine how many points from large $r$ toward the peaks
 to be included is to make $\chi^2/d.o.f$ as close to 1 as possible; the
 $\chi^2/d.o.f$'s for the eight $T$'s are $1.16$,$1.96$,$1.17$,$0.46$,$0.82$,$0.16$,$1.16$,$1.23$ for
 monopole-antimonopole correlators and  $3.65$,$2.87$,
 $2.16$,$2.48$,$2.02$,$1.06$,$0.80$,$1.36$ for monopole-monopole correlators. The values of $\frac{1}{R_d \cdot T}
 \equiv \frac{M_d}{T}$
 obtained are all in $2-3$ region with slight tendency to decrease
 toward higher $T$.

The resulting values of magnetic coupling from fitting both
monopole-antimonopole and monopole-monopole correlators for all
available temperatures are shown in
Fig.\ref{fig_running_coupling}(a): as expected $\alpha_M$ is
getting weaker at low $T$ and stronger at high $T$. Inverting
these values (because of the Dirac condition (\ref{Dirac})) one
gets the respective electric couplings, which we compare to the
dashed green curve corresponding to the pure gauge SU(2) one-loop
asymptotic freedom expression $\alpha_{E} \approx
\frac{2\pi}{(22/3)Ln[C*(T/T_c)]}$ with the coefficient $C=54$
determined from the last point at $T=3.8T_c$. The one-loop running
becomes much slower as $T$ decreases, which is normal and shall be
cured by higher order loops and more importantly by
non-perturbative corrections such as instanton, see
\cite{Randall_Shuryak}.

Finally, in Fig.\ref{fig_running_coupling}(b) we combine
 the data on the magnetic coupling and  the monopole density from the same work
 to get the dimensionless magnetic plasma parameter (\ref{plasma_gamma})
 in the studied range $T\approx 1-4T_c$. We use a formula $n(T)/T^3=0.557/[Ln(2.69*T/T_c)]^2$ from
\cite{D'Alessandro:2007su} which was found to nicely fit the
monopole density data. The $\alpha_M$ used is the average of the
fitted values from monopole-antimonopole and monopole-monopole
correlator, with half of their difference as the error bar. As one
can see, magnetic component of QGP never gets to be a weakly
coupled gas, as $\Gamma_M>1$ at all $T$ even close to $T_c$. On
the other hand, even at the highest $T$ the value of $\Gamma_M$
does not reach large values $>10$ at which liquids are known to
become glass-like and viscosity starts growing.

In the  high $T$ limit we expect ``magnetic
scaling'' $n_M \sim (\alpha_E T)^3$, thus
plasma parameter approach a fixed point $\Gamma_M \sim
\alpha_M n_M^{1/3}/T\rightarrow \Gamma_M^*$ due to
 Dirac condition.
The curve in Fig.3(b)  indicate
$\Gamma_M^* \sim 5$.

\section{Summary}
In this Letter we have shown that gauge theory monopoles
in a deconfined phase behave as charges in a Coulomb plasmas.
Furthermore, we show that
 the temperature dependence (running)
of the magnetic couplings in gauge theories is indeed
the inverse of the electric one,
electric-magnetic duality
arguments \cite{Liao_ES_mono}.
 Good agreement was found (in shape and magnitude) of the
correlators we calculated in MD simulation  \cite{Liao_ES_mono} for
novel electric-magnetic plasmas and recent lattice results
\cite{D'Alessandro:2007su}.
More specifically,
 we concluded that the magnetic part of QGP at $T=(1-4)T_c$
has an effective plasma parameter in the  ``good liquid'' domain \cite{Comment}
$\Gamma=1-4.5$,  not spoiling the
``perfect liquid'' observed at RHIC. We predict existence of a fixed
point for $\Gamma$ at high $T$.\\

{\bf Acknowledgments.} \vskip .2cm We are grateful to
 A.~D'Alessandro and M.~D'Elia,
who provided us with extensive set of their
data in table form, going beyond the content of their
 paper \cite{D'Alessandro:2007su}. We thank Claudia Ratti for helpful
 discussions. This work was supported in parts by the US-DOE grant
DE-FG-88ER40388.

\end{document}